\documentclass[a4paper]{jpconf}

\usepackage{graphicx}
\usepackage[numbers,sort&compress,square]{natbib}
\usepackage{amsmath}  	
\usepackage{xcolor}
\usepackage{siunitx}

\begin{document}

\title{Hardware-in-the-loop wind-tunnel testing of wake interactions between two floating wind turbines}

\author{Alessandro Fontanella, Kristjan Milic, Alan Facchinetti, Sara Muggiasca and Marco Belloli}

\address{Dept. of Mechanical Engineering, Politecnico di Milano, via La Masa 1, 20156 Milano, Italy}

\ead{alessandro.fontanella@polimi.it}

\begin{abstract}
Wake interactions in floating wind farms are inherently coupled to platform motion, yet most experimental studies to date neglect this two-way coupling by prescribing platform movements. This work presents a hardware-in-the-loop (HIL) wind-tunnel methodology to investigate wake interactions between two floating wind turbines with fully coupled aerodynamic loading and platform dynamics. The approach integrates physical wind-tunnel testing of two scaled rotors with a real-time numerical model that accounts for platform motion, mooring restoring forces, and hydrodynamic loads. Experiments conducted under low-turbulence inflow conditions show that a downstream turbine operating in the wake of an upstream turbine experiences reduced mean thrust and platform deflections due to the decreased inflow velocity, alongside enhanced low-frequency platform motions driven by increased turbulent energy in the wake. The proposed HIL framework provides a controlled experimental basis for studying wake-induced excitation mechanisms and supports the validation of floating wind farm models and control strategies.
\end{abstract}

\vspace{1em}

\section{Introduction}
Floating wind technology is rapidly scaling up, and the industry now faces the challenge of moving from small demonstration projects, typically involving one or a few turbines, to commercial-scale floating wind farms. This presents challenges that extend well beyond individual turbine performance and encompass multidisciplinary aspects such as the aerodynamic interactions among turbines within an array. Wake interactions are already known to play a central role in bottom-fixed wind farms, where they reduce overall energy capture, accelerate turbine fatigue damage, and drive the need for farm-level operational strategies \cite{Meyers2022}.

Recent research has shown that the wakes of floating wind turbines differ substantially from those of bottom-fixed wind turbines, primarily due to the larger platform motions enabled by their compliant foundations. Under low-turbulence inflow conditions, platform motions imprint distinct periodic structures in the wake, with frequencies matching those of the dominant platform oscillations \cite{Fontanella2025,Messmer2025}. These unsteady wake features give rise to time-varying inflow velocities that can excite downstream turbines operating within the wake \cite{Fontanella2025draft}. At the same time, motion-induced wake dynamics may enhance wake recovery, an effect that is particularly pronounced under low ambient turbulence, where wakes are otherwise more persistent \cite{Messmer2024,Messmer2025,Fontanella2025}. In addition, floating wind turbines typically operate at larger rotor tilt angles due to platform inclination, resulting in an upward deflection of the wake \cite{Carmo2024}. These mechanisms occur simultaneously and are intrinsically coupled to the dynamic response of the turbines.

The problem of aerodynamic interactions in floating wind farms is inherently two-way coupled: platform motions influence wake development, and the resulting wake in turn governs the aerodynamic loading and motion response of the turbines operating within it. Because few full-scale floating arrays currently exist, wind tunnel experiments offer a crucial means of studying these coupled interactions in controlled conditions. However, most existing scale-model studies have treated the downstream turbine using prescribed or fixed motions \cite{Messmer2024,Messmer2025,Fontanella2025}, thereby neglecting the two-way coupling between wake-induced loads and platform response. This simplification limits the physical realism of experimental results.

In this work, we address this limitation by developing a hardware-in-the-loop (HIL) methodology that enables dynamically coupled wind-tunnel testing of two floating wind turbines. HIL approaches have been used in recent years to study the response of individual floating turbines \cite{Taruffidraft,Fontanella2023b}, but they have not been applied to turbine-to-turbine interactions. Here, the method is extended for the first time to capture the full aero-hydro-servo-elastic coupling within a floating wind farm. 
In this sense, this work does not aim to provide fully scaled quantitative load predictions for floating wind farms. Instead, it targets the experimental reproduction and validation of two-way coupled wake–motion dynamics under controlled conditions, enabling investigation of mechanisms that cannot currently be observed at full scale.

\section{Methodology}
The floating wind farm replicated by means of the HIL approach is composed of two wind turbines aligned with the wind direction at a spacing of 5.75$D$. The floating wind turbines are based on the DTU 10 MW reference wind turbine \cite{Bak2013} mounted on top of the TripleSpar platform, a hybrid semisubmersible–spar floater with catenary mooring \cite{Lemmer2020}.
The floating wind farm was recreated in the atmospheric boundary layer test section of the Politecnico di Milano wind tunnel, measuring \SI{13.84}{\meter} (width) $\times$ \SI{3.84}{\meter} (height) $\times$ \SI{35}{\meter} (length). The end of the test section is equipped with a turning table of \SI{13}{\meter} diameter that hosts the experiment.

\subsection{Wind farm scaling and implications for the HIL approach}
The floating wind turbines are modeled at a geometric scale of 1:150, selected based on the ratio between the DTU 10~MW rotor diameter and the dimensions of the wind-tunnel test section, allowing turbine spacings representative of commercial wind farms (5--10 rotor diameters).

The velocity scale is set to 1:2.5, independently of the geometric scale and without enforcing Froude-number similitude, in order to avoid excessively low model-scale wind speeds and thereby preserve inflow quality. With this choice, the model-scale rated wind speed is \SI{4.6}{\meter\per\second}, whereas Froude scaling would result in unrealistically low velocities, with rated conditions around \SI{1}{\meter\per\second}.

The combined length and velocity scales yield a time scale of 1:60 and an acceleration scale of 24:1. Consequently, gravity-driven and inertial effects cannot be physically scaled, and the weight and inertial loads of the floating wind turbine are not correctly reproduced in the physical model.
Given these limitations, the HIL approach focuses on reproducing the inflow and aerodynamic loads generated by the rotor, while hydrodynamic and rigid-body dynamics are provided by the numerical model. Aerodynamic loads are reconstructed from tower-top force measurements, which include contributions from rotor aerodynamics as well as inertia and gravity effects; the inertia- and gravity-related components are compensated for within the HIL control loop.

\subsection{Aerodynamic representativeness for wake--motion coupling}
Each turbine has a \SI{1.2}{\meter} diameter rotor. The selected velocity scale ensures acceptable inflow quality at model scale, although strict Reynolds-number similarity is not achieved. To mitigate this limitation, the rotor blades are redesigned to reproduce the thrust coefficient and the spanwise distribution of normal aerodynamic loads of the full-scale turbine, as well as their sensitivity to angle-of-attack variations induced, for example, by platform motion \cite{Fontanella2023}.

This strategy ensures that the integral thrust and its low-frequency variations are accurately reproduced. Since thrust is the primary aerodynamic driver of platform surge and pitch motions, the resulting rigid-body response to aerodynamic loading is preserved within the HIL framework. Moreover, thrust magnitude and distribution over the rotor largely govern the strength and spatial structure of the wake velocity deficit, so that the main features of wake formation and propagation are also retained.

Under these conditions, the experiment reproduces coupled aerodynamic--structural mechanisms responsible for wake-induced excitation of downstream turbines. Absolute load magnitudes and motion amplitudes are not fully reproduced due to rotor redesign approximations and simplified hydrodynamic modeling. Nevertheless, the dominant physical processes underlying wake--motion coupling are consistently captured.

\subsection{Experimental setup and HIL system architecture}
\begin{figure}[tb]
\begin{center}
\includegraphics[width=\textwidth]{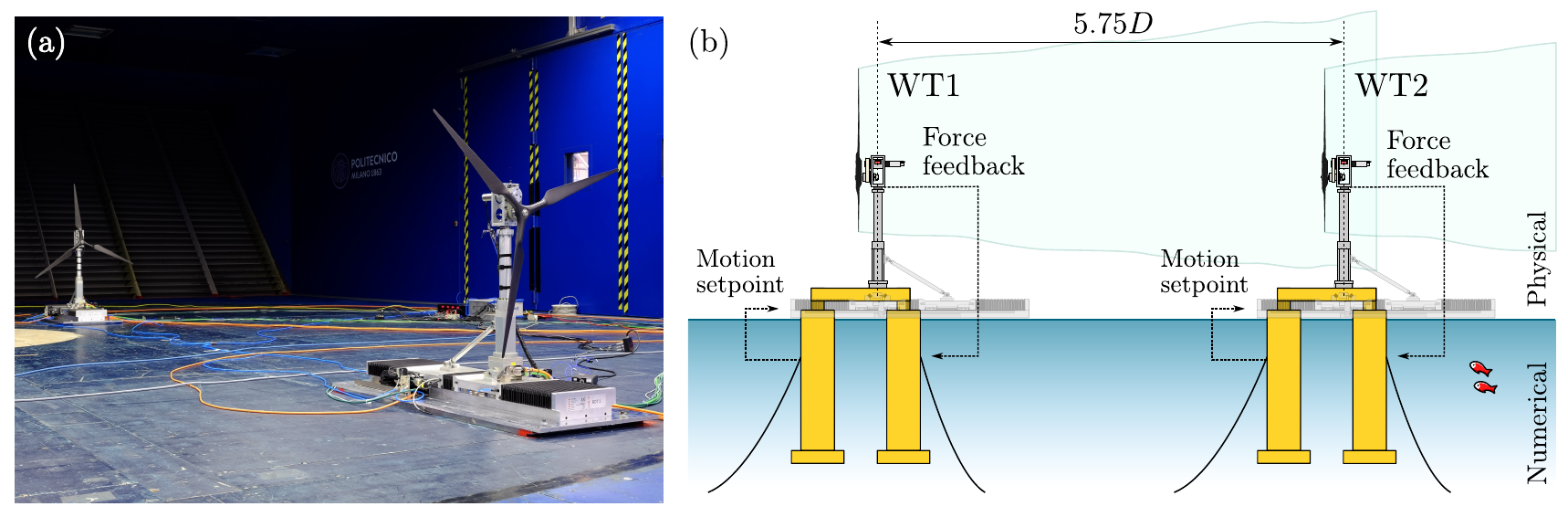}
\end{center}
\vspace{-0.5cm}
\caption{(a) Experimental setup of the floating wind farm in the wind tunnel. (b) Schematic representation of the hardware-in-the-loop system used to emulate the dynamic behavior of the floating wind farm.}
\label{fig:Setup}
\end{figure}

The experimental setup and its integration in the HIL framework are shown in Fig.~\ref{fig:Setup}, and the key properties of the simulated floating wind turbines are summarized in Table~\ref{tbl:properties}.
The rotors were mounted on a robotic system with a parallel kinematic structure that replicates platform surge and pitch motions \cite{Fontanella2024}.
\begin{table}[tb]
\caption{Key properties of the floating wind turbines used in the wind tunnel experiments.}
\begin{center}
\begin{tabular}{lccc}
\hline
Parameter 							& Model scale & Full scale \\ \hline
Rotor diameter ($D$) $[$m$]$				& 1.20 	       & 178.4 \\
Hub height ($h_\mathrm{hub}$) $[$m$]$						& 0.84	       & 124.1\\
Tower base height $[$m$]$				& 0.84	       & 8.8\\
Platform mass $[$kg$]$					& 17.04	       & $5.6\cdot10^7$\\
Rotor-nacelle mass (physical model) $[$kg$]$	& 3.31	       & $1.1\cdot10^7$\\
Rotor-nacelle mass (numerical model) $[$kg$]$	& 0.20	       & $6.8\cdot10^6$\\
Platform surge frequency $[$Hz$]$  			& 0.30	       & 0.005\\
Platform pitch frequency $[$Hz$]$   			& 2.40	       & 0.040\\
\hline
\end{tabular}
\end{center}
\label{tbl:properties}
\end{table}

The HIL methodology divides the floating wind farm into two subsystems. 
The physical subsystem consists of the wind-tunnel inflow and the scaled rotors, which reproduce the aerodynamic interactions within the wind farm. 
The numerical subsystem models the rigid-body dynamics of the two floating turbines, including mooring-line restoring forces, and the hydrodynamic loads associated with waves propagating through the farm.
The two subsystems are coupled through the HIL control strategy. The controller reconstructs the aerodynamic loads generated by the rotors and passes them to a numerical rigid-body dynamics model of the floating wind turbines, in which they are combined with simulated hydrodynamic forces. The numerical model then computes the resulting surge and pitch motions, which are prescribed in real time to the robotic platforms.

\subsection{Real-time HIL control and dynamic modeling}
\begin{figure}[tb]
\begin{center}
\includegraphics[width=0.8\textwidth]{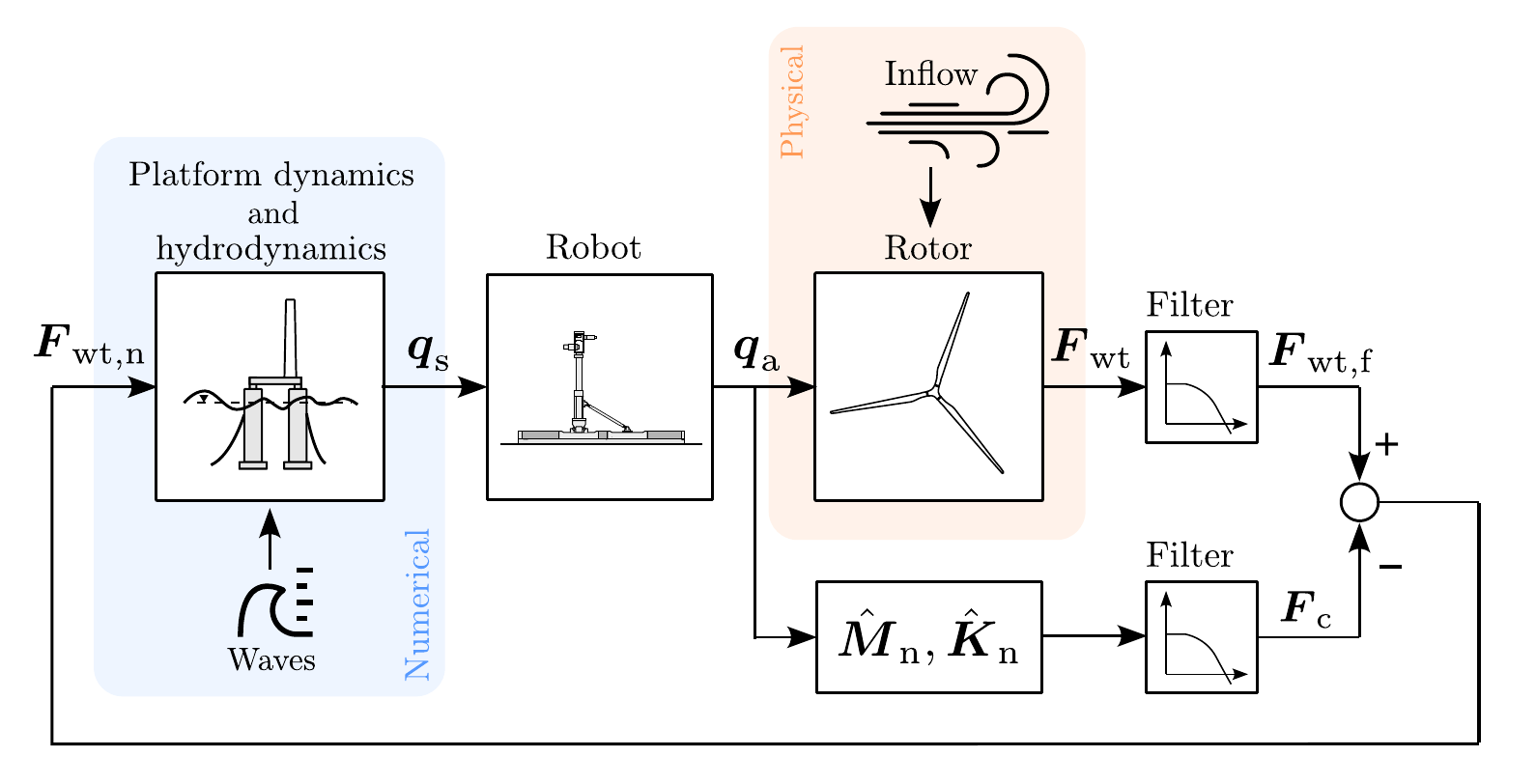}
\end{center}
\vspace{-0.5cm}
\caption{Block diagram of the hardware-in-the-loop control framework.}
\label{fig:HILcontrol}
\end{figure}

The HIL control framework is illustrated in Fig.~\ref{fig:HILcontrol}.
The HIL controller operates in parallel on the two floating wind turbines. For each turbine, the surge and pitch equations of motion given in Eq.~\ref{eq:EOM} are solved in real time on a dedicated machine using a model-scale time step of \SI{1}{\milli\second}. The equations of motion are expressed as
\begin{equation}
(\boldsymbol{M}_\mathrm{fowt}+\boldsymbol{A}\infty)\ddot{\boldsymbol{q}}_\mathrm{s} + \boldsymbol{R}_\mathrm{hydro}\dot{\boldsymbol{q}}_\mathrm{s} + \boldsymbol{K}_\mathrm{hs,moor}{\boldsymbol{q}}_\mathrm{s} = \boldsymbol{F}_\mathrm{wave} + \boldsymbol{F}_\mathrm{wt,n}
, ,
\label{eq:EOM}
\end{equation}
where $\boldsymbol{q}_\mathrm{s}=[x_\mathrm{s},\beta_\mathrm{s}]'$ is the vector of platform surge and pitch motions obtained from numerical integration of the equations of motion. The matrix $\boldsymbol{M}_\mathrm{fowt}$ represents the mass matrix of the floating wind turbine, including contributions from the tower and the rotor–nacelle assembly, while $\boldsymbol{A}_\infty$ is the infinite-frequency added-mass matrix. The matrix $\boldsymbol{R}_\mathrm{hydro}$ accounts for hydrodynamic damping, and $\boldsymbol{K}_\mathrm{hs,moor}$ is the stiffness matrix associated with hydrostatics, mooring system, and gravity. The vector $\boldsymbol{F}_\mathrm{wave}$ represents wave-induced forces, whereas $\boldsymbol{F}_\mathrm{wt,n}$ denotes the filtered aerodynamic loads generated by the rotor and expressed in the platform reference frame.

The matrix $\boldsymbol{M}_\mathrm{fowt}$ is derived from a rigid-body model of the floating wind turbine, composed of tower, nacelle, hub, and blades. Each component is characterized by its mass, center of gravity, and inertia tensor in a local reference frame, and its contribution is projected onto the surge and pitch degrees of freedom. In the stiffness matrix $\boldsymbol{K}_\mathrm{hs,moor}$, the mooring stiffness contribution is obtained from the floating platform specifications \cite{Lemmer2020}, the gravitational contribution is derived from the rigid-body model of the turbine, and the hydrostatic contribution is computed using a panel-based potential-flow model of the floating platform. The same panel-based approach is employed to evaluate the added-mass matrix $\boldsymbol{A}_\infty$, while hydrodynamic damping is represented through the matrix $\boldsymbol{R}_\mathrm{hydro}$, which accounts, within a linearized formulation, for the combined effect of all sources of hydrodynamic damping acting on the platform.

The hydrodynamic model used in the HIL framework prioritizes computational efficiency while retaining sufficient fidelity for wake-interaction studies. Linear potential-flow theory is employed to represent hydrostatic stiffness, added mass, and wave-induced loads, while higher-order wave–structure interactions, viscous nonlinearities, and radiation memory effects are neglected. This formulation captures the dominant dynamic response characteristics of the floating turbines and enables stable real-time integration within the HIL control loop. The framework can be readily extended to higher-fidelity hydrodynamic models when wave-induced effects become the primary focus.

\subsection{Aerodynamic load reconstruction and force feedback}
The aerodynamic loads are obtained by subtracting the inertia and gravity loads $\boldsymbol{F}_\mathrm{c}$ from the filtered force measured at the tower top by the load cell, which in the present setup is an ATI Mini45 sensor with SI-580-20 calibration:
\begin{equation}
\boldsymbol{F}_\mathrm{wt,n} = \boldsymbol{F}_\mathrm{wt,f}-\boldsymbol{F}_\mathrm{c}
\, .
\label{eq:ForceFB}
\end{equation}
The vector $\boldsymbol{F}\mathrm{c}$ represents the inertia and gravity loads of the rotor–nacelle assembly induced by surge and pitch motions and is expressed as:
\begin{equation}
\boldsymbol{F}_\mathrm{c} = {\boldsymbol{M}}_\mathrm{n}\ddot{{\boldsymbol{{q}}}}_\mathrm{a} + {\boldsymbol{K}}_\mathrm{n}{{\boldsymbol{{q}}}}_\mathrm{a}
\, ,
\label{eq:ForceFeedback}
\end{equation}
where $\boldsymbol{q}_\mathrm{a} = [x_\mathrm{a},\beta_\mathrm{a}]'$ is the vector of the actual surge and pitch motions of the robotic platforms, measured by linear encoders integrated into the robots. The inertia and gravity loads are assumed to be proportional to the rigid-body acceleration of the nacelle through the matrix ${\boldsymbol{M}}_\mathrm{n}$ and to its displacement through the matrix ${\boldsymbol{K}}_\mathrm{n}$. The matrices ${\boldsymbol{M}}\mathrm{n}$ and ${\boldsymbol{K}}\mathrm{n}$ are identified from prescribed-motion tests in still air using system identification techniques, such that the resulting ${\boldsymbol{F}}_\mathrm{c}$ minimize the aerodynamic force $\boldsymbol{F}_\mathrm{wt,n}$ under these conditions.

\section{Results}
The HIL system is first verified through tests conducted in the absence of wind, which assess the ability to reproduce the surge and pitch response of the floating wind turbines as represented in the numerical model. Following this verification phase, the HIL framework is applied under wind conditions to reproduce the turbine response to the inflow developing within the wind-tunnel floating wind farm configuration, and the measured response is related to the measured inflow conditions.

All quantities reported in this section are given at full scale.

\subsection{Verification tests with no wind}
In the no-wind condition, the force measured by the tower-top load cell consists solely of inertia and gravity contributions. These tests are therefore used to verify that the proposed HIL control strategy effectively cancels such contributions and that the resulting residual forces are sufficiently small so as not to significantly affect the dynamics of the floating wind turbine simulated by the HIL system.

To assess the capability of the HIL system to reproduce the rigid-body surge and pitch response of the floating wind turbine, free-decay tests are performed. In these tests, a fictitious wave force
\[
\boldsymbol{F}_\mathrm{wave} = [1, h_\mathrm{hub}]' \cdot F_{x,\mathrm{step}}
\]
is applied to the system, where $F_{x,\mathrm{step}}$ is a step force that transitions from \SI{-1100}{\kilo\newton} to zero at the start of the decay test. As a result of this force, the platform reaches an initial displacement of approximately \SI{-12}{\meter} in surge and \SI{-2}{\degree} in pitch. Once the force is removed, the platform undergoes free oscillations in both degrees of freedom.

The same test is repeated with $\boldsymbol{F}_\mathrm{wt,n}$ set to zero (``open loop'') in order to measure the response of the floating wind turbine without the effect of the HIL force feedback and to provide a reference for comparison.

\begin{figure}[tb]
\begin{center}
\includegraphics[width=\textwidth]{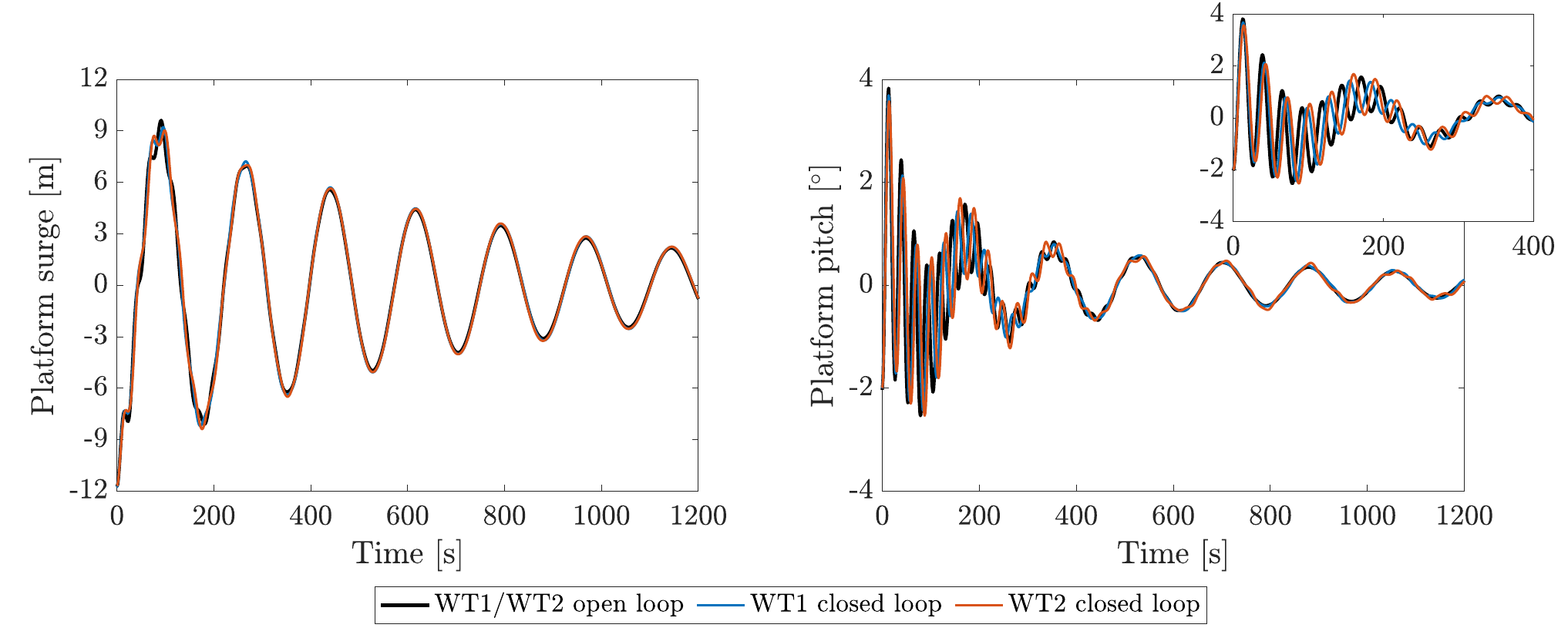}
\end{center}
\vspace{-0.5cm}
\caption{Platform surge and pitch motions of the two wind turbines from free-decay tests conducted in still air, without wind-turbine force feedback (``open loop'') and with force feedback (``closed loop'').}
\label{fig:DecayNowind}
\end{figure}

The response to the force step is shown in Fig.~\ref{fig:DecayNowind}. In both surge and pitch, the response exhibits oscillations arising from the coupled dynamics of the two modes. This coupling is particularly evident in the pitch response, where higher-frequency, lower-amplitude oscillations associated with the pitch mode are superimposed on lower-frequency oscillations corresponding to the surge mode, especially during the initial phase of the decay.

Overall, the response obtained with the force feedback active (``closed loop'') closely matches the response obtained without force feedback, demonstrating that the aerodynamic force estimation method identifies the inertia and gravity contributions associated with surge and pitch motions with sufficient accuracy. The surge response with closed-loop control reproduces the open-loop response more accurately than the pitch response. The low-frequency oscillations associated with the surge mode show very good agreement between the two configurations in both surge and pitch and are consistent across the two floating wind turbines.

Larger discrepancies are observed at the higher-frequency oscillations associated with the pitch mode, which are more evident in the pitch response. In particular, the closed-loop response exhibits a slightly lower pitch frequency compared to the open-loop case, and differences are observed between the two floating wind turbines. These discrepancies are attributed to residual inertia and gravity load contributions at the platform pitch frequency and to differences in the residuals between the two robotic platforms.

\subsection{Steady wind test}
In the steady wind test, the free-stream inflow is spatially uniform and nearly laminar, with a turbulence intensity of approximately 2\%. This condition was selected because the effect of platform motion on wake development is known to be more pronounced under low-turbulence inflow conditions. The inflow velocity is constant and equal to \SI{12.3}{\meter\per\second} at full scale. In the numerical model, the water is still, and no waves are present, allowing the isolation of aerodynamic effects and wake interactions in the response of the floating wind turbines. The wind turbines are operated at fixed rotor speeds which are kept constant throughout the tests.

The rotor speed of WT1 is set to 9.5~rpm, corresponding to a tip-speed ratio of 7.3, close to the rotor optimal operating point. Under these conditions, WT1 develops a thrust force of \SI{1841}{\kilo\newton}, corresponding to a thrust coefficient of $C_t = 0.82$. This thrust level defines the strength of the wake generated by WT1.

\begin{figure}[tb]
\begin{center}
\includegraphics[width=\textwidth]{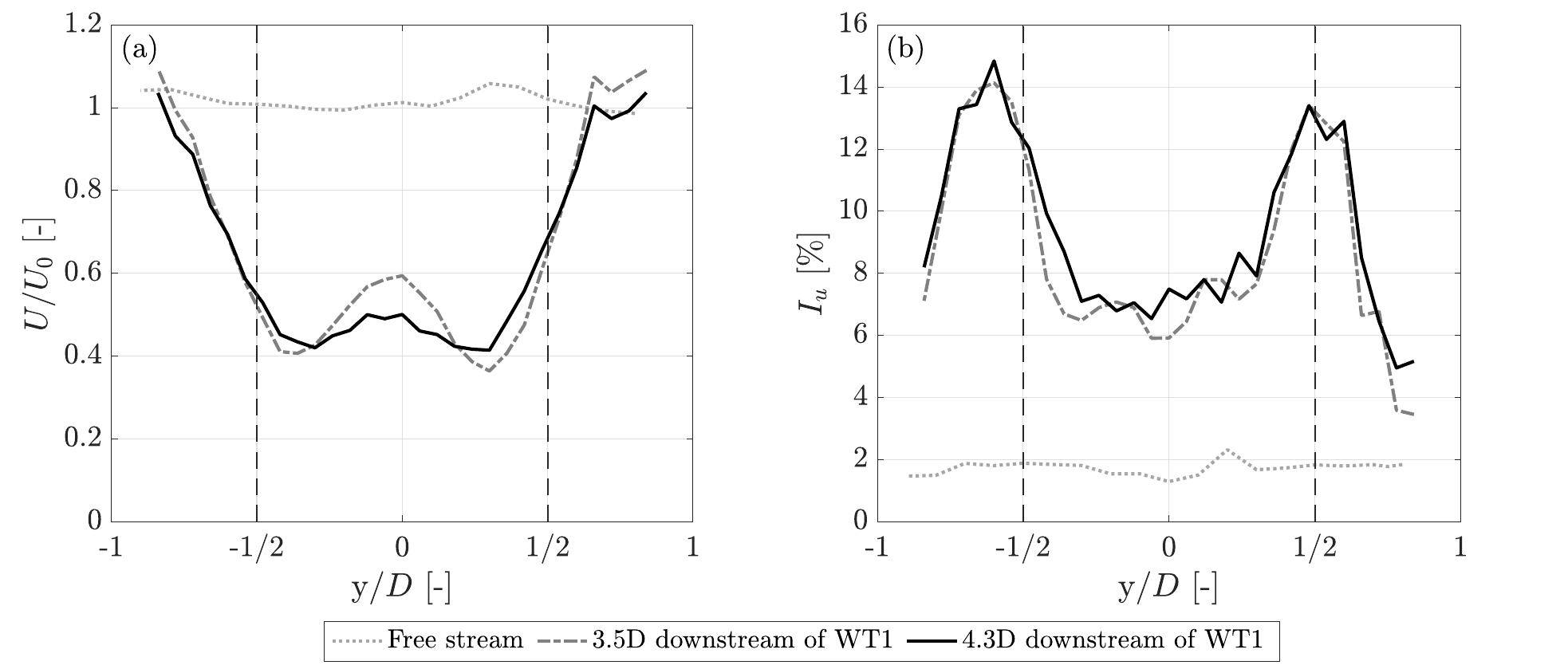}
\end{center}
\vspace{-0.5cm}
\caption{Inflow conditions along a horizontal line at hub height, shown for the free-stream region and two locations between WT1 and WT2. \textbf{(a)} Mean wind speed normalized by the free-stream value. \textbf{(b)} Turbulence intensity.}
\label{fig:FlowField}
\end{figure}

The wake generated by WT1 is measured at two downstream locations by means of hot-wire probes, at distances of 3.5$D$ and 4.3$D$ from the rotor. Figure~\ref{fig:FlowField} shows the time-averaged wind speed and turbulence intensity along horizontal lines at hub height, measured 1.3$D$ upstream of WT1 (free stream) and at 3.5$D$ and 4.3$D$ downstream. These measurements characterize the wake development of WT1 and define the inflow conditions experienced by WT2.
At 3.5$D$ downstream, the average velocity profile exhibits a double-Gaussian shape, while at 4.3$D$ a gradual transition towards a top-hat profile is observed. The turbulence intensity is significantly higher than in the free stream, reaching approximately 7\% at the center of the wake and increasing to about 14\% near the wake edges.

WT2 operates fully immersed in the wake generated by WT1. To adapt its operation to the reduced inflow velocity, the rotor speed of WT2 is set to 6~rpm. The combined effect of reduced inflow velocity and reduced rotor speed results in a thrust force of \SI{627}{\kilo\newton}, corresponding to approximately 34\% of the thrust force developed by WT1.

\begin{figure}[tb]
\begin{center}
\includegraphics[width=\textwidth]{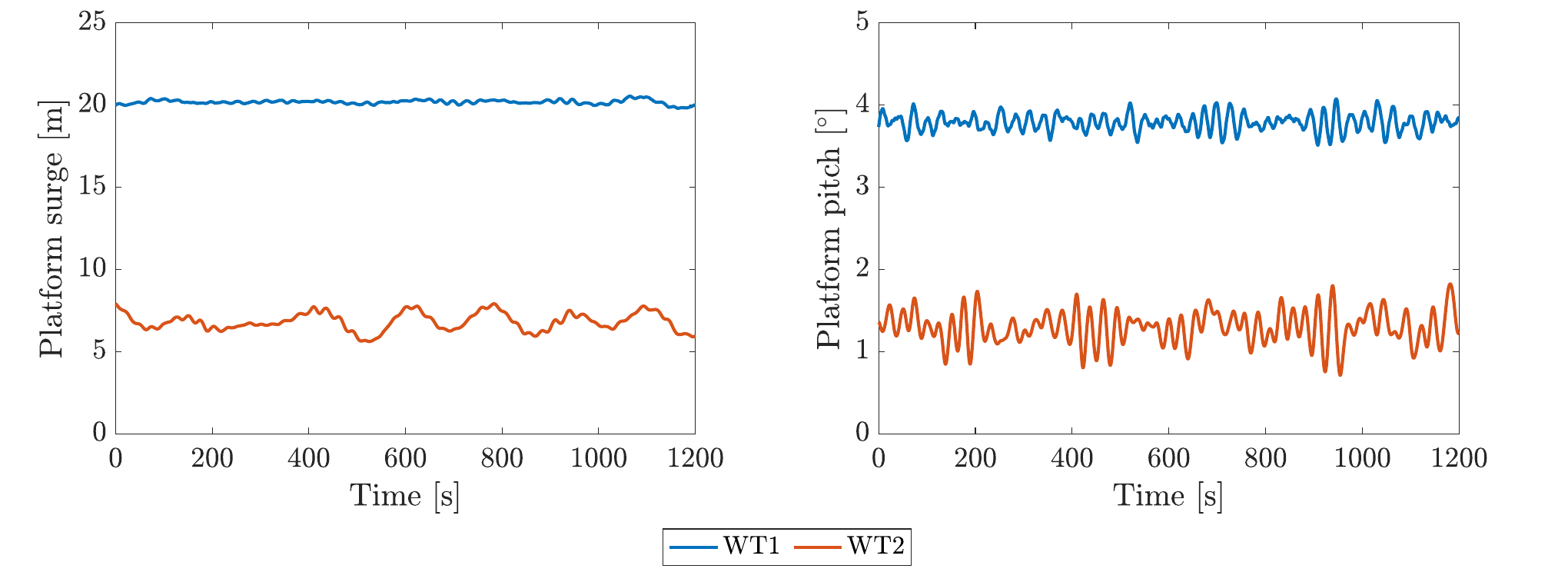}
\end{center}
\vspace{-0.5cm}
\caption{Platform surge and pitch motions of the two wind turbines with wind.}
\label{fig:SteadyState_B1B2}
\end{figure}

The platform response under these inflow conditions is computed by the HIL control system and reproduced in the wind tunnel by the robotic platforms. Figure~\ref{fig:SteadyState_B1B2} shows the time series of surge and pitch motions of the two floating wind turbines under wind loading. The time-averaged platform displacements in both degrees of freedom are consistent with the thrust forces developed by the two turbines. WT1 exhibits a mean surge displacement of \SI{20.1}{\meter} and a pitch rotation of \SI{3.8}{\degree}. For WT2, the mean surge and pitch displacements are \SI{6.81}{\meter} and \SI{1.3}{\degree}, respectively, corresponding to approximately 34\% of those of WT1. This is consistent with the reduction in aerodynamic thrust force acting on WT2, given the identical characteristics of the floating platforms.

\begin{figure}[tb]
\begin{center}
\includegraphics[width=\textwidth]{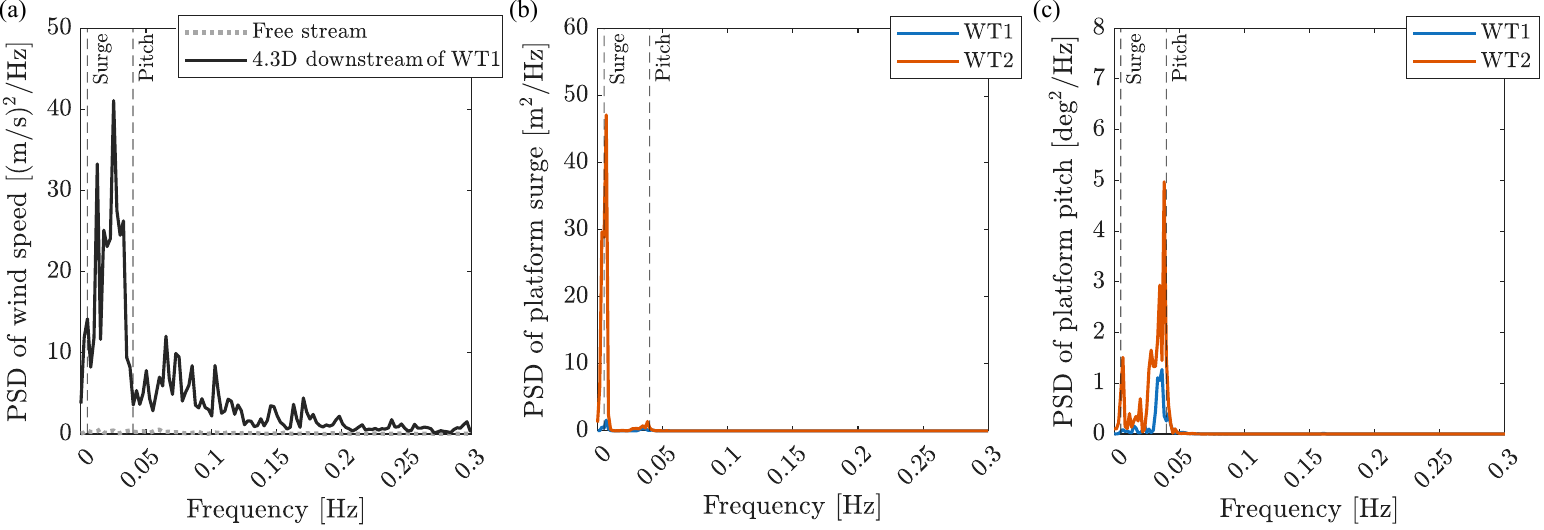}
\end{center}
\vspace{-0.5cm}
\caption{Power spectral density (PSD) of inflow and platform response.
\textbf{(a)} PSD of the inflow velocity measured upstream of WT1 and in the wake of WT1 at 4.3$D$ downstream of the rotor, at hub height and at a lateral offset equal to half a rotor diameter.
\textbf{(b)} PSD of the platform surge motion of WT1 and WT2.
\textbf{(c)} PSD of the platform pitch motion of WT1 and WT2.
In all panels, vertical dashed lines indicate the natural frequencies of the platform surge and pitch modes.}
\label{fig:SteadyState_B1B2ws_PSD}
\end{figure}

In addition to the reduction in mean response, the motions of WT2 are characterized by higher variability compared to WT1. This behavior is attributed to the increased turbulence intensity of the inflow in which WT2 operates, which enhances the excitation of low-frequency platform motions. This link is investigated in Fig.~\ref{fig:SteadyState_B1B2ws_PSD}, which shows the power spectral density (PSD) of the inflow for WT1 and WT2 together with the PSD of the platform response.
The PSD of the inflow is obtained from velocity measurements at hub height, at a lateral offset equal to half a rotor diameter, at 1.3$D$ upstream of WT1 (free stream) and 4.3$D$ downstream in the wake. Compared to the free-stream inflow, the wake exhibits increased turbulent energy at low frequencies, particularly in correspondence of the natural frequencies of the platform surge and pitch modes. This increased low-frequency energy content is reflected in the response of WT2, whose platform motions exhibit amplified peaks at the platform natural frequencies compared to WT1, which operates in the low-energy free-stream inflow.

\section{Conclusions}
This work presents a hardware-in-the-loop wind tunnel methodology to investigate wake interactions between two floating wind turbines with two-way coupling between wake-induced aerodynamic loading and platform motion. The approach combines physical wind-tunnel testing of two rotors with a real-time numerical model that provides platform dynamics driven by aerodynamic loads, hydrostatic and mooring restoring effects, and hydrodynamic forces.

The HIL framework was first verified through free-decay tests in still air. These tests demonstrated that the adopted aerodynamic force-reconstruction strategy effectively compensates inertia and gravity contributions in the tower-top load cell measurements, resulting in surge and pitch responses that closely match the reference provided by the numerical model. Minor discrepancies were observed at the pitch natural frequency and are attributed to residual force components and small differences between the robotic platforms.

Under wind conditions with low free-stream turbulence intensity, the upstream turbine generated a strong wake characterized by a pronounced velocity deficit and increased turbulence levels at 3.5$D$--4.3$D$ downstream. When operating fully immersed in this wake, the downstream turbine experienced a reduced mean thrust and correspondingly lower mean surge and pitch deflections. In addition, its motion response exhibited increased low-frequency variability and amplified spectral peaks at the platform natural frequencies. Analysis of the inflow spectra confirmed that the wake contains higher low-frequency energy relative to the free stream, explaining the increased excitation of the downstream floating turbine.

This first experimental application demonstrates the practical value of HIL wind-tunnel testing by enabling controlled and repeatable investigation of coupled wake–motion interactions. This capability is particularly relevant given the scarcity of full-scale measurements of floating wind farm wakes and the challenges in validating high-fidelity multi-turbine simulations. Although the adopted scaling strategy does not reproduce absolute load magnitudes or fully scaled motion amplitudes, it preserves the integral aerodynamic loading and platform dynamic characteristics governing wake development and wake-induced motions.
Accordingly, the HIL approach provides an experimental basis for validating aero–hydro–servo-dynamic models, identifying and quantifying wake-induced excitation mechanisms relevant to fatigue and control, and supporting the development and assessment of floating wind farm layout and control strategies.

Future work will extend the experimental campaign to a broader range of inflow conditions and sea states, including controlled variations in turbine spacing and alignment and the inclusion of wave-induced forcing, in order to investigate the combined influence of atmospheric turbulence, platform dynamics, and wave excitation on wake interactions in floating wind farms.

\ack
This study was carried out within the NEST – Network 4 Energy Sustainable Transition (D.D. 1243 02/08/2022, PE00000021) and received funding under the National Recovery and Resilience Plan (NRRP), Mission 4, Component 2, Investment 1.3, funded from the European Union-NextGenerationEU. This manuscript reflects only the authors' views and opinions, neither the European Union nor the European Commission can be considered responsible for them.

\bibliographystyle{iopart-num}
\bibliography{Torque26_references}

\end{document}